\documentclass[structabstract]{aa}
\usepackage{graphicx}
\usepackage{natbib}
\usepackage{bm}
\usepackage{txfonts}
\begin{document}
   \title{Turbulent resistivity evaluation in MRI generated turbulence.}


   \author{G. Lesur
          \inst{1}
          \and
          P-Y. Longaretti\inst{2}
          }
   \institute{Department of Applied Mathematics and Theoretical Physics, University of Cambridge, Centre for Mathematical Sciences,
Wilberforce Road, Cambridge CB3 0WA, UK
           \and Laboratoire d'Astrophysique, UJF CNRS, BP 53, 38041 Grenoble Cedex 9, France\\
              \email{g.lesur@damtp.cam.ac.uk\\
}
                     }

   \date{Received XXX, year; accepted XXX}


  \abstract
   {MRI turbulence is a leading mechanism for the generation of an efficient turbulent transport of
   angular momentum in an accretion disk through a turbulent viscosity effect. It is believed that the
   same process could also transport large-scale magnetic fields in disks, reshaping the magnetic
   structures in these objects. This process, known as turbulent resistivity, has been suggested and
   used in several accretion-ejection models and simulations to produce jets. Still, the efficiency
   of MRI-driven turbulence to transport large-scale magnetic fields is largely unknown. }
   {We present new analytical and numerical results aiming at quantifying the turbulent resistivity
   produced by MRI-driven turbulence in accretion disks.}
   {We investigate this problem  both analytically and numerically. We introduce a linear calculation
   of the MRI in the presence of a spatially inhomogeneous mean magnetic field. We show that, in this
   configuration, MRI modes lead to an efficient magnetic field transport, on the order of the angular
   momentum transport. We next use fully non linear simulations of MRI turbulence to compute the turbulent
   resistivity in several magnetic configurations.}
   {We find that the turbulent resistivity is on the order of the turbulent viscosity in all our simulations,
   although somewhat lower. The variations in the turbulent resistivity are correlated with the variation in
   the turbulent viscosity as a function of the imposed mean field. Finally, the turbulent resistivity tensor
   is found to be highly anisotropic with a diffusion coefficient 3 times greater in the radial direction than
   in the vertical direction.}
   {These results support the possibility of driving jets from turbulent disks; the resulting
   jets may not be steady.}


   \maketitle
%

\section{Introduction}

After several decades of debate, and in spite of significant
progress, the origin and efficiency of angular momentum transport
remains a central problem in accretion disks physics. Turbulence is
one of the major physical mechanism through which the
observationally well-constrained anomalous transport can be
achieved; indeed, the first $\alpha$ model \citep{SS73} already
assumed a strong level of turbulence, leading to an effective
viscosity orders of magnitude higher than molecular viscosity.
However, the physical origin of this turbulence in disks is still
largely discussed.

Various sources of turbulence, both hydrodynamical and
magneto-hydrodynamical, have been proposed.
Subcritical turbulence (\citealt{RZ99} and references therein), if
present, is too inefficient \citep{LL05,JBSG06}. Convection is also too
inefficient and transports angular momentum in the wrong direction
\citep{C96,SB96}. Two-dimensional turbulence driving by
small-scale, incoherent gravitational instabilities \citep{G96} or
baroclinic instabilities \citep{KB03} is a possible option,
although the last one is still highly controversial
\citep{JG06,PSJ07}.

In a seminal paper, \cite{BH91a} indentify an MHD
instability, the magnetorotational instability (MRI) that drives
turbulence in the nonlinear regime. This instability provides the
most extensively studied transport mechanism, mainly with the help
of local unstratified \citep{HGB95} and stratified \citep{SHGB96}
3D simulations, and global \citep{H00} disk simulations. These
simulations have shown that MRI turbulence is an efficient way to
transport angular momentum, although the role of microphysical
processes has largely been underestimated \citep{LL07,FPLH07}.

MRI turbulence may also produce resistive transport\footnote{In
this paper, a transport process refers to any physical
mechanism by which a magnetic field line is moved from one place
to another, without assuming anything about flux-freezing
conditions.} (transport of magnetic fields through a ``turbulent
resistivity" process) in disks. This transport is a key ingredient
of accretion-ejection models (see, e.g., \citealt{F97,CF00} and
references therein). The related turbulent resistivity $\eta_T$ is
parameterized with the Shakura-Sunyaev ansatz as
$\eta_T=\alpha_{\eta} V_A H$ ($V_A$ is the Alfv\'en speed based on
the mean field); stationary accretion ejection models require an
anisotropy of the turbulent resistivity transport of few; furthermore, 
a very efficient turbulent transport with
$\alpha_\eta \lesssim 1$ is required to produce stationary
structures. Less efficient transport would not prevent the
launching of jets from disks, but these structures would then be
nonstationary. The question of whether turbulent resistivity is
an adequate source of field line slippage through the gas to
launch a jet is still controversial in the literature, and is
further discussed in the concluding section of this paper.

In any case, turbulent resistivity is an issue in its own right,
and in this paper, we present new numerical results aimed at
quantifying more precisely the resistive transport due to MRI
turbulence. Our approach differs from the method recently
introduced by Brandenburg, R\"adler, Schrinner and coworkers to
compute the turbulent resistivity tensor (see \citealt{SRSRC05},
\citealt{BRRK08} and references therein). Their method relies on
an expansion of the mean electromotive force with respect to the
mean field and its derivatives, and makes use of appropriately
chosen test field, whose evolution is computed along with the
turbulent flow. Our approach is also different from the one used
by \cite{GG09}, in which a magnetic structure is imposed as an
initial condition, and the resistive properties of the background
turbulent flow are deduced from the decay time of this structure.
The method we adopt here is described in section 2, and the
physical content of the two methods is discussed there. More
generally, section 2 describes the physics and the numerical
methods we have used to study turbulent resistivity in disks.
Section 3 presents a linear analysis of the model, which sheds
some light on the numerical findings described in section 4.
Section 5 discusses the astrophysical implications of these
results along with some future line of work.

\section{Shearing-box equations and numerical method}

MRI-related turbulence has been extensively studied in the
literature, but little attention has been devoted to the question
of turbulent-driven resistivity yet. One of the major advances of
the recent years has been the realization that an accurate
determination of turbulent transport properties requires an
accurate representation of all scales down to the dissipation
scales \citep{LL07,FPLH07}, although this accuracy has not yet
been achieved for astrophysically relevant magnetic Prandtl and
Reynolds numbers regimes. We do not imply that actual
astrophysical Reynolds numbers need to be resolved in simulations
to obtain reliable results, an obviously hopeless task anyway;
but for large enough Reynolds and magnetic Reynolds numbers, one
expects the dissipation scale to be decoupled from the transport
scales, in which case one could use a closure model without
resolving the dissipation scales. However, today simulations don't
seem to have reached this regime yet.

Shearing box simulations offer a particularly convenient setting
to quantify turbulent resistivity; however, the boundary
conditions prevent the existence of the large scale gradients of
mean magnetic field required for characterizing the transport of
this mean field produced by turbulence. We bypass this difficulty
by a prescription whose physical motivation and formulation will
be described later on in this section. For the time being, we
briefly recall the basic equations for the shearing-box model,
which has been largely studied and used in the literature. The
reader may consult \cite{HGB95}, \cite{B03} and \cite{RU08} for an
extensive discussion of the properties and limitations of this
model.

Since MHD turbulence in disks is essentially subsonic, we will
work in the incompressible approximation, which allows us to
eliminate sound waves and density waves from the problem. 
Although density waves are excited in shearing box turbulence
\citep{HP08b} they should not have a big
impact on the turbulence spectrum itself or on turbulent transport
as their amplitude decays exponentially fast when one goes to
small scales \citep{HP08a}. This has also been confirmed by direct
comparison between incompressible and compressible simulations
\citep{FPLH07}. We also neglect vertical stratification,
consistently with the local shearing-box model \citep{RU08}.
Explicit molecular viscosity and resistivity are included in our
description.

The shearing box equations follow from a local approximation. We
chose a Cartesian box centered at $r=R_0$, rotating with the disk
at angular velocity $\Omega=\Omega(R_0)$ and having dimensions
$(L_x,L_y,L_z)$ with $L_i \ll R_0$. Assuming $R_0\phi \rightarrow
x$ and $r-R_0 \rightarrow -y$, one eventually obtains the shearing
box equations:
\begin{eqnarray}
\nonumber \partial_t \bm{U}+\bm{\nabla\cdot} (\bm{U\otimes U})&=&-\bm{\nabla} \Pi+\bm{\nabla\cdot }(\bm{ B \otimes B}) \\
& &-2\bm{\Omega \times U}+2\Omega S y \bm{e_y}+\nu \bm{\Delta U},\\
\partial_t \bm{B}&=&\bm{\nabla \times} (\bm{U \times B}) +\eta \bm{\Delta B},\\
\label{divv} \bm{\nabla \cdot U}&=&0,\\
\bm{\nabla \cdot B}&=&0.
\end{eqnarray}
The boundary conditions associated with this system are periodic
in the $x$ and $z$ direction and shearing-periodic in the $y$
direction \citep{HGB95}. In these equations, we have defined the
mean shear $S=-r\partial_r \Omega$, which is set to
$S=(3/2)\Omega$, assuming a Keplerian rotation profile. The
generalized pressure term $\Pi$ includes both the gas pressure
term $P/\rho_0$ and the magnetic one $\bm{B}^2/2\rho_0$. This
generalized pressure $\Pi$ is actually a Lagrange multiplier
enforcing equation (\ref{divv}), and is therefore computed by
solving a Poisson equation. Note also that the magnetic field is
expressed in Alfv\' en-speed units, for simplicity.

The steady-state solution to these equations is the local
Keplerian profile $\bm{U}=Sy\bm{e_x}$. In this paper, we will
consider only the turbulent deviations from this Keplerian
profile.  These may be written as $\bm{V}=\bm{U}-Sy\bm{e_x}$,
leading to the following equations for $\bm{V}$:

\begin{eqnarray}
\nonumber \partial _t \bm{V}+\bm{\nabla \cdot}(\bm{V \otimes
V})&=&-\bm{\nabla} \Pi+\bm{\nabla\cdot }
(\bm{ B \otimes B})-Sy\partial_x \bm{V}\\
\label{motion}& & +(2\Omega-S) V_y\bm{e_x}-2\Omega V_x \bm{e_y}+\nu\bm{\Delta V},\\
 \nonumber \partial _t \bm{B}&=&-Sy\partial_x \bm{B}+S B_y\bm{e_x}\\
 \label{induction}& & +\bm{\nabla \times} (\bm{V \times B}) +\eta \bm{\Delta B},\\
\label{Vstruct} \bm{\nabla \cdot V}&=&0,\\
\label{Bstruct}\bm{\nabla \cdot B}&=&0.
\end{eqnarray}

Following \cite{HGB95}, one can integrate the induction equation
(\ref{induction}) over the volume of the box, leading to:

\begin{equation}
\label{consB}
\frac{\partial \langle \bm{B}\rangle }{\partial
t}=S\langle B_y\rangle \bm{e}_x,
\end{equation}

\noindent where $\langle \rangle$ denotes a volume average.
Therefore, the mean magnetic field is conserved, provided that no
mean radial field is present. In this work, the mean field will be
either vertical or azimuthal.

To numerically solve the shearing-box equations, we use a spectral
Galerkin representation of equations
(\ref{motion})--(\ref{Bstruct}) in the sheared frame
\citep[see][]{LL05}. This frame allows us to use a Fourier
decomposition since the shearing-sheet boundary conditions are
transformed into perfectly periodic boundary conditions. Moreover,
this decomposition allows us to conserve magnetic flux to machine
precision without any modification, which is an advantage compared
to finite-difference or finite-volume methods (the total magnetic
flux created during one simulation is typically $10^{-11}$).
Equations (\ref{Vstruct}) and (\ref{Bstruct}) are enforced to
machine precision using a spectral projection \citep{P02}. The
nonlinear terms are computed with a pseudospectral method, and
aliasing is prevented using the 3/2 rule. To always compute the
physically relevant scales in the sheared frame, we use a remap
method similar to the one described by \cite{UR04}. This routine
redefines the sheared frame every $T_{\mathrm{remap}}=L_x/(L_yS)$
and we have checked that none of the behaviour we describe in this
paper was related to this numerical timescale. Since spectral
methods are very little dissipative by nature, we check that
numerical dissipation is kept to very small values, computing the
total energy budget at each time step \citep[see][for a discussion
of this procedure]{LL05}. We therefore ensure that numerical
dissipation is responsible for less than $3\%$ of the total
dissipative losses occurring in these simulations.

To quantify the dissipation processes in the
simulations, we use dimensionless numbers defined as:
\begin{itemize}
\item The Reynolds number, $Re=SL_z^2/\nu$, comparing the nonlinear advection term to the viscous dissipation.
\item The magnetic Reynolds number, $Rm=SL_z^2/\eta$, comparing magnetic field advection to the Ohmic resistivity.
\item The magnetic Prandtl number, defined as the ratio of the two previous quantities $Pm=Rm/Re=\nu/\eta$, which
measures the relative importance of the dissipation processes,
and, correlatively, is related to the ratio of the viscous and
resistive dissipation scales.
\end{itemize}
In the following we use $S^{-1}$ as the unit of time and $SL_z$ as
the unit of velocity. One orbit corresponds to
$T_\mathrm{orb}=3\pi S^{-1}$. For simplicity, we keep the same
notation for dimensionless and dimensional quantities.

\subsection{Turbulent resistivity definition}

Our aim is to test to which extent the effect of MRI turbulence on
the mean field can be modelled as a turbulent resistivity on large
scales, and characterize inasmuch as possible the resulting
turbulent resistivity tensor. We therefore distinguish the large
scale mean field $\overline{\bm{B}}$ and velocity
$\overline{\bm{V}}$ and the fluctuating (turbulent) fields
$\bm{b}$ and ${v}$. We assume $\langle \bm{b}\rangle=0$ and
$\langle \bm{v} \rangle=0$ where $\langle\rangle$ denotes an
ensemble (or time, under the ergodic hypothesis) average. The
induction equation reads:

\begin{equation}
\label{meanB}
\partial_t{\bm{\overline{B}}}=\bm{\nabla \times (\overline{V}\times \overline{B})}+\nabla\times\bm{\mathcal{E}}
\end{equation}

\noindent where we have defined the mean electromotive force (EMF)
$\bm{\mathcal{E}}=\langle{\bm{v}\times
\bm{b}}\rangle$. The turbulent resistivity hypothesis
assumes:

\begin{equation}
\label{EMFexp}
\mathcal{E}_i=\beta_{ijk}\partial_{x_k}\overline{B}_j.
\end{equation}

\noindent Our main objective is to test and quantify this
assumption. We keep no term proportional to $\overline{B}$
($\alpha$-like effect), as we found out that they are not produced
in our simulations.

At this point, two different routes are open to study the
turbulent diffusion of the magnetic field.

\begin{itemize}
  \item In the first one, Eq.~(\ref{EMFexp}) is \textit{assumed}.
  One can split Eq.~(\ref{induction}) into an equation for the mean field
  $\overline{\bm B}$ and one for the deviation from the mean $\bm{b}$.
  As the induction equation is linear in the field, $\bm b$
  formally depends on the mean and turbulent velocity fields and mean
  magnetic field only. One
  then introduces extraneous magnetic fields $\bm b^{pq}$, whose equation of
  evolution is the same as for $\bm{b}$, except for the mean field $\overline{B}$ which is replaced
  by a \emph{adhoc} test field $\overline{B}^{pq}$. It is then possible to derive the EMF associated
  to this test field $\bm{\mathcal{E}}^{pq}=\langle \bm{v} \times \bm{b^{pq}}\rangle$.

  One can deduce some of the components of $\beta_{ijk}$ using several test fields and computing
  the correlations between $\overline{B^{pq}}$ and $\mathcal{E}^{pq}$. By construction, this method
  probes the velocity field $\bm{v}$ with a test field $\bm{B^{pq}}$ which is \emph{different} from
  the real field $\bm{B}$ entering in the Lorentz force. Therefore, one assumes implicitly
  that the properties of $\beta_{ijk}$ don't depend on the topological properties of $\bm{B}$, which might not be true
  for subcritical dynamos \citep[see e.g.][]{LO08a} or turbulence driven through the Lorentz force like MRI turbulence.
  This method is detailed in \cite{SRSRC05} and \cite{BRRK08} and references therein.

  \item Alternatively, one can \textit{impose} the constancy of some large
  scale component of the magnetic field gradient throughout
  the evolution and test if Eq.~(\ref{EMFexp}) accurately represents
  the effect of turbulence on field diffusion. As in the previous approach,
  adequate choices of the imposed field configuration allow us to characterize various
  elements of the diffusivity tensor. This approach is adopted here and specified
  in more detail in the following subsection.
\end{itemize}

These two approaches do not quite probe the same physical aspects
of the problem. This first one is justified inasmuch as the mean
field hypothesis Eq.~(\ref{EMFexp}) holds, provided that the
turbulent resistivity tensor does no depend on the field topology.
The second one is justified without assumption if the turbulent
diffusion of the field is much slower than the other time-scales
associated to the turbulence dynamics. However, even if this
requirement is not satisfied, the physical content of this second
approach can be grasped by comparing the questions of magnetic
field and momentum transfer. Consider for example a simple
unmagnetized plane Couette flow. With shearing box boundary
conditions, the turbulent flow is unable to react of the shear
(velocity gradient) imposed on the flow through the boundary
conditions. As shown by \cite{LL05}, the flow nevertheless becomes
turbulent above some transition Reynolds number $R_c$ and produces
a shearwise transport characterized by $R_c$. Alternatively, one
can let the flow react on the imposed velocity gradient by
adopting e.g.\ wall boundary conditions in the shearwise
direction. This changes has little effect on the transition
Reynolds numbers (at least for cyclonic rotation: see
\citealt{LL05}), and does not appear to affect much the
dimensionless transport, as suggested by the numerical results of
\cite{LL05} and the experimental results compiled in
\cite{DDDLRZ05}. These considerations make it plausible that the
turbulent diffusion of the magnetic field with an imposed field
gradient or with a self-consistently evolving one should be
similar, although a more systematic study is required to ascertain
this statement.

In practice, we chose field configurations such that only one
component of the field gradient is non vanishing at any given
time, so that the current is directly proportional to this
non-vanishing gradient. In this situation, one can dispense with
the third order tensor and use

\begin{equation}
\mathcal{E}_i=-\eta_{ij}J_j
\end{equation}

\noindent instead, where $\bm{J}=\bm{\nabla \times \overline{B}}$ is the mean
electric current and $\eta$ is a constant tensor, but the reader
should keep in mind that there is no reason that $\eta_{ijk}$ should
be antisymmetric in the relevant indices. Similarly, $\eta$ has no
reason to be diagonal for MRI turbulence. In this work, we explore
this model looking for linear correlations between the components
$\mathcal{E}_j$ and $J_i$, and computing the related
resistivity tensor coefficients.

To keep in line with the logic of the $\alpha$ parameterization as
adapted to an incompressible fluid (see \citealt{LL07}), one may
introduce an $\alpha$ tensor relating to the turbulent $\eta$
tensor:

\begin{equation}\label{alphaeta}
  \alpha_{\eta_{ij}}\simeq \frac{\eta_{ij}}{SL_z^2}.
\end{equation}

\noindent With our choice of units ($S=1$ and $L_z=1$), one has
$\alpha_{\eta_{ij}} = \eta_{ij}$, and the distinction between the
two quantities is usually dropped in the remainder of the paper.

\subsection{Numerical protocol}

To compute the turbulent resistivity, one needs a mean current,
which is not present in local (shearing box) simulations. We
obtain this current by imposing a large scale and \emph{non
homogeneous} field in the box. This is done with the help of our
Galerkin representation by forcing the largest Fourier mode in one
direction to a constant value $\Delta B_0$. To trigger the MRI, we
impose a mean field $B_0$ which can be either azimuthal or
vertical. The aspect ratio is $L_x\times L_y\times
L_z$=$4\times2\times1$ where $x=\phi$, $y=-r$ and $z=z$. The
factor 2 in $L_y$ allow us to trigger more easily secondary
instabilities in the strong mean vertical field cases (see
\citealt{GX94} and \citealt{BMCRF08}). The resolution used is
$128\times 128 \times 64$, similar in cell size to the one used in
\cite{LL07}. The Reynolds number is kept constant at $Re=1600$ as
well as the magnetic Prandtl number, which is fixed at unity:
$Pm=1$. This corresponds to a Elsasser number $\Lambda\equiv
V_{A}^2/\eta \Omega=24$ for $B_0=0.1$ and $\Lambda=1.5$ for
$B_0=0.025$. Each simulation is integrated over 500 shear times
(to obtain meaningful time averages), and the average are computed
from 400 last shear times (to avoid initial conditions artifacts).

To postprocess the results, we first compute the time average of
$\bm{B}$ and $\mathcal{\bm{E}}_j$. We then use a script which
extracts the mean current and compute the correlation with the
emfs, giving in the end one component of the $\eta$ tensor for one
run. Note that using this procedure, we can compute resistivity
associated with $B_z(y)$ (radially varying vertical field),
$B_x(y)$ (radially varying azimuthal field) and $B_x(z)$
(vertically varying azimuthal field) configurations. Imposing a
constant radial field $B_y$ would lead to a linear growth of the
azimuthal field $B_x$, which may not have any physical meaning. We
cannot compute turbulent resistivity associated with non
axisymmetric structures because of the shearing-sheet boundary
conditions.

The runs we have performed do not test all the dependencies of the
dimensionless turbulent resistive transport with respect to all
the dimensionless parameters of the problem, but only a subset of
them, namely:

\begin{itemize}
\item The role of the mean field orientation $B_0$, and of the
orientation the imposed field varying component $\Delta B_0$.
\item The role of the relative amplitude of the mean and imposed
varying field components $\epsilon= \Delta B_0/B_0$. In the
process, only a few components of the turbulent resistivity tensor
are probed.
\item The role of the relative mean field amplitude,
characterized by\footnote{Our $\beta$ parameter is akin to the
usual plasma beta parameter, as both differ only by a factor of
order unity in a stratified disk in vertical hydrostatic
equilibrium.} $\beta= (4\pi \rho S^2 H^2)/B_0^2=B_0^{-2}$ (the
last equality follows from our choice of units).
\end{itemize}

No characterization with respect to the dissipation parameters has
been attempted (most notably the magnetic Prandtl number), a
problem requiring substantial efforts and that will be addressed
elsewhere.

Our local simulation box physical size is on the order of the
scale height at most. As such, one would expect to mimic the
effect of large scale gradients of the magnetic field by small
values of $\epsilon$. However sizeable (but of limited extent)
local field gradients may also be present in disks as a result of
the dynamics, and it is of some interest to explore values of
$\epsilon$ of order unity as well.

\section{Analytic results}

It is of some interest to look into the linear behavior of the
instability first, as it does provide some insight into the
turbulent problem that we discuss in the next section. Indeed,
there is evidence that the channel mode (both a linear and
nonlinear solution to the incompressible equations) plays some
role in the overall transport properties of the MRI-driven
turbulent states.

For this reason, we want to test the turbulent resistivity
hypothesis (equations \ref{meanB} and \ref{EMFexp}) in the linear
regime. This is done assuming the fluctuations $v$ and $b$ are
infinitely small compared to the averaged fields. We then
calculate $v$ and $b$ by a classical linear analysis and deduce
the EMF $\bm{\mathcal{E}}$ which is a quadratic quantity in this
limit\footnote{For this reason, this approach is also called
"quasi-linear".} . The same procedure can be used for Maxwell and
Reynolds stresses, allowing one to compute ``effective" transport
coefficients in the linear regime.

\begin{figure}
   \centering
   \includegraphics[scale=0.4]{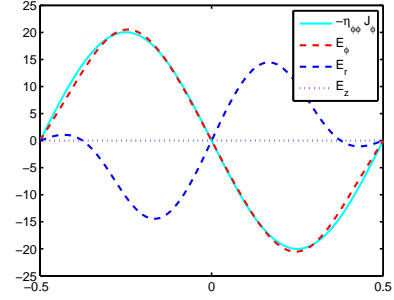}
   \caption{Mean emfs and best fit with the current for $\beta=100$
   and $\epsilon=0.3$ for the fastest growing channel mode with an imposed vertical mean field and
   radially sinusoidally varying imposed azimuthal current. The relevant emf component $E_\phi$ is seen to
   behave as a resistive term. This component $E_\phi$ measures the diffusion of the vertical field in the
   radial direction. The amplitudes are arbitrary.}
              \label{transpchan}%
\end{figure}

To make comparisons easier, we investigate one of the configurations which is
simulated in the next section of this paper, namely we impose a
vertical mean field of the form

\begin{equation}\label{eqB}
B_T(y)= B_0\left[1+\epsilon\cos(k_0 y)\right],
\end{equation}

\noindent where $k_0=2\pi/L$ ($=2\pi$ in our choice of units) and
$\epsilon$ measures the relative amplitude of the sinusoidal
component of the mean field, and is to be taken as a small
expansion parameter later on in some of the analytic expressions.
Other configurations would lead to a similar qualitative behavior,
which is the point of interest here.

\begin{figure}
   \centering
   \includegraphics[scale=0.4]{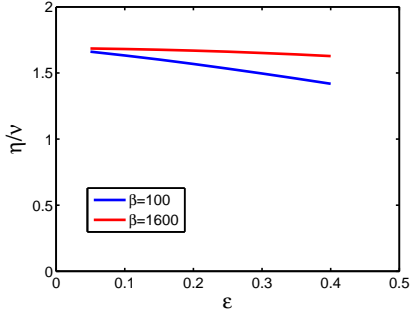}
   \caption{Ratio of the mean emf $E_\phi$ and mean Reynolds stress tensor ($r,\phi$)
   component due to the most unstable channel mode.}
              \label{etanu}%
\end{figure}

\begin{figure*}
   \centering
   \includegraphics[scale=0.5]{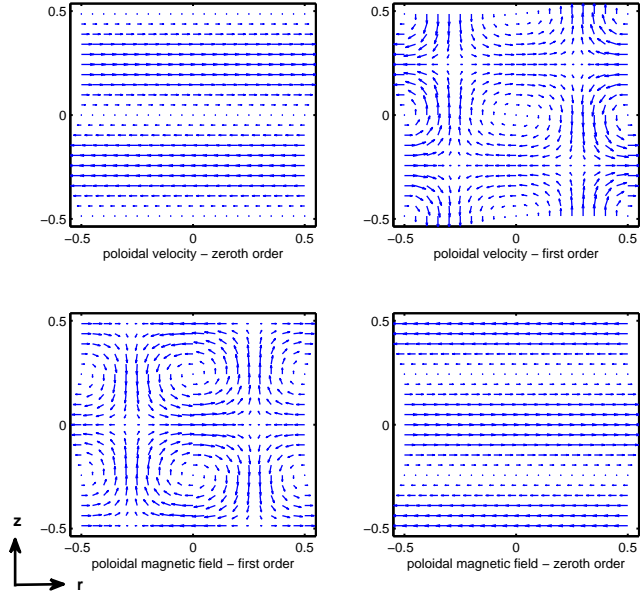}
   \caption{Zeroth and first order components of the poloidal magnetic and velocity fields for the fastest growing channel
   mode ($\beta=100$, $\epsilon=0.3$). The coupling of
   zeroth (resp.\ first) order component of the velocity field to the first (resp.\ zeroth) order component
   of the magnetic field gives rise to the transport of magnetic field by the channel mode.}
              \label{ubcorr}%
\end{figure*}

Assuming that the development of the instability is fast compared
to the viscous and resistive time-scales, we neglect the
corresponding terms in Eqs.~(\ref{motion}) and (\ref{induction})
in this linear analysis. Under this approximation, the system
Eq.~(\ref{motion}) to (\ref{Vstruct}) admits a simple equilibrium
with constant pressure and varying equilibrium azimuthal
velocity\footnote{An actual disk is indeed expected to react to
the presence of a superimposed magnetic pressure gradient at some
initial time by first adjusting its velocity profile instead of
its gas pressure profile. Nevertheless, we have also explored the
converse situation, where the magnetic pressure variation is
balanced by a variation of the gas pressure, without modification
of the Keplerian profile (a behavior expected in an incompressible
fluid). This change of equilibrium has little effect on the
results reported in this section.} profile to balance the magnetic
pressure gradient:

\begin{equation}\label{eqV}
  V(y) = Sy -\frac{1}{4\Omega_0}\frac{d B_T^2}{dy}.
\end{equation}

This leads us to introduce the total shear $S_T$:

\begin{equation}\label{ST}
  S_T=S-\frac{1}{4\Omega_0}\frac{d^2 B_T^2}{dy^2}.
\end{equation}

It turns out that only channel-like mode present correlations of
the EMF that behave as expected under the resistivity hypothesis,
so we specialize from the onset to that type of modes. First order
linear deviations from the mean stationary solution described
above are referred to as $\bm{U}_1$, $\bm{B}_1$, $P_1$, and one
considers only axisymmetric perturbations of the form

\begin{equation}\label{perturb}
  \bm{X}_1(y,z,t)=\bm{x}(y)\exp(i\omega t - i k z),
\end{equation}

\noindent where $\bm{X}$, $\bm{x}$ refers to the velocity,
magnetic, and gaz pressure fields.

It is useful to distinguish the poloidal and toroidal components
of the velocity and magnetic fields:

\begin{eqnarray}
  \bm{v} & = & \bm{v}_\perp + v_x\bm{e}_x, \label{pol}\\
  \bm{b} & = & \bm{b}_\perp + b_x\bm{e}_x.\label{tor}
\end{eqnarray}

Because, of the assumed axisymmetry, on can introduce potential
fields $\psi$ and $\varphi$ for these poloidal components such
that

\begin{eqnarray}
  \bm{v}_\perp & = & i k \psi
  \bm{e}_y+\frac{d\psi}{dy}\bm{e}_z,\label{potv}\\
  \bm{b}_\perp & = & i k \varphi
  \bm{e}_y+\frac{d\varphi}{dy}\bm{e}_z,\label{potb}
\end{eqnarray}

All variables can be expressed in terms of $\psi$, which leads to
a second order ODE for $\psi$. One obtains:

\begin{eqnarray}\label{psi}
  (\omega^2 & - & V_A^2k_z^2)\psi'' - 2 V_b^2k_z^3\psi'\nonumber \\
    & + & \left[(\omega^2
  -V_A^2k_z^2)-\frac{\kappa_T^2 \omega^2+2\Omega_0 S_T V_A^2 k_z^2}{\omega^2
  -V_A^2k_z^2}\right] k_z^2 \psi = 0,
\end{eqnarray}

\noindent where $V_A(y)$ and $V_b(y)$ are local Alfv\'en speed
type quantities, and $\kappa_T(y)$ is a local epicyclic frequency,
defined by

\begin{eqnarray}
  V_A^2 & = & B_T^2,\label{VA}\\
  V_b^2 & = & \frac{B_T B_t'}{k_z},\label{Vb}\\
  \kappa_T^2 & = & 2\omega_0(2\Omega_0-S_T).\label{KT}
\end{eqnarray}

The magnetic field $\bm{b}$ and the azimuthal velocity $v_x$ can
be expressed in terms of $\psi$ as follows:

\begin{eqnarray}
  b_x & = & - \frac{k_z B_T}{\omega}v_x - \frac{k_z^2 S_T
  B_T}{\omega^2}\psi,\label{bx}\\
  b_y & = & -i \frac{k_z^2 B_T}{\omega}\psi,\label{by}\\
  b_z & = & -\frac{k_z B_T}{\omega}\psi'-\frac{k_z
  B_T'}{\omega}\psi,\label{bz}\\
  v_x & = & -k_z\psi\frac{\omega^2(S_T-2\Omega_0)-S_T V_A^2
  k_z^2}{\omega(\omega^2-V_A^2 k_z^2)}.\label{vx}
\end{eqnarray}

We have solved Eq.~(\ref{psi}) numerically with periodic radial
boundary conditions\footnote{No boundary condition is implied
vertically so that the modes are not discretized in this
direction. In this section, the unit of length is set to $L_y$,
but this has no incidence on the results or on the rest of the
paper.}. Analytic solutions can be found in terms of series
expansion in $\epsilon$ in the form
$\psi=\psi_0+\psi_1+\psi_2\ldots$ and
$\omega^2=\omega_0^2+\omega_1^2+\omega_2^2\ldots$; the first order
solution is rather straightforward to derive, but the explicit
form of this solution is not particularly illuminating and will
not be given in detail, except for some relevant features that
will be mentioned whenever needed.

Solutions in the $\epsilon=0$ limit are the usual MRI modes, i.e.
simple sinusoidal modes $\psi_0\propto \exp(\pm i k_y y)$.
Solutions for arbitrary $\epsilon$ generalize these modes. It
turns out that only the nodeless mode ($k_y=0$) produces a mean
emf that correlates with the imposed current\footnote{The reason
for this can be understood by looking at the equation for the
first order correction $\psi_1$, which shows that the correct
radial behavior for coupling occurs only in this case.}, and in
the following, only these modes are studied.

For definiteness, we show results pertaining to the fastest
growing mode, for $\beta=100$ and $\epsilon=0.3$. Similar results
are obtained for other sets of parameters, and for other purely
vertical modes.

Following the quasi-linear analysis procedure, we define the
quadratic emf as:

\begin{equation}
\langle\bm{\mathcal{E}}(y)\rangle=\frac{1}{2}\Re \Big[ \bm{v \times b^\dagger} \Big]
\end{equation}

\noindent where $\langle \rangle$ denotes in this case a vertical
average of the fluctuations, and the dagger stands for complex
conjugation. As the channel mode is exponentially growing in time,
the exponential prefactor has been left out (or equivalently, a
time-dependent $\mathcal{E}$ would be required in the above
equation). The various components of the emf due to the unstable
mode, as well as the best fit of Eq.~(\ref{EMFexp}) with the
imposed current are shown on Fig.~\ref{transpchan}.

It is of some interest to compare the efficiency of the
``resistive" and ``viscous" transport due to the instability,
i.e., the value of $\eta_{\phi\phi}=emf/current$ deduced here to
the value of $\nu_{r\phi}=\nu=(Reynolds\ +\ Maxwell\
stress)/shear$ due to the same mode. This is shown on
Fig.~\ref{etanu} for the same mode and for two different values of
$\beta$ and a range of values of $\epsilon$. Note that this ratio
is constant throughout the growth of the mode. The analytic result
for this ratio in the limit of small $\epsilon$ is
$\eta_{\phi\phi}/\nu_{r\phi}=27/16\simeq 1.7$. Apparently, the
relative efficiency of resistive and viscous transports due to the
instability are of similar magnitude, more or less independently
of the amplitude of the imposed field variation and mean field
strength.

The comparable efficiency of the two transport effects due to the
channel mode comes somewhat as a surprise. From a thermodynamical
point of view, the instability sets in because of the imposed
velocity shear and tries to restore a thermodynamical equilibrium
by transporting momentum to reduce this shear. At first sight,
there is no reason that it should also try to suppress a gradient
in magnetic field with about the same efficiency.

However, if one considers the second-order correlation approximation
used in mean field electrodynamic, one finds \citep[e.g.][]{RR07}:
\begin{equation}
\beta_{ij} \sim \frac{1}{3}\langle\bm{v}^2\rangle \tau,
\end{equation}
where $\tau$ is a typical correlation timescale of the small scale
turbulence. This kind of result can be understood as a mixing
length theory for magnetic fields involving velocities of the
order of $v$ and a length on the order of $\tau v$. Although the
second-order correlation approximation is not entirely justified
in the quasi-linear theory used here, at least some of the terms
of the induction equation provide a mixing length behavior. This
suggests that with a typical timescale on the order of the shear
time, the dimensionless turbulent resistivity should be of the
order of the turbulent transport (the turbulent transport being of
the order of the turbulent kinetic energy). Pushing this argument
one step further, we note that in general in numerical
simulations, $\langle b_x b_y\rangle\sim3\langle v_x v_y\rangle$,
implying that one should observe $\beta_{ij}\sim\alpha/4$. This is
not consistent with the analytic result, but it agrees very well
with the numerical results of the next section for a vertical
field and field variation; however it does not capture the origin
of the significant anisotropy of the resistivity tensor observed
in these simulations.

\smallskip

The analytic results as well as the preceding argument imply that
the motion and magnetic field deviations due to the varying field
should be coupled. An illustration of the efficiency of these
couplings leading to this high resistive transport can be obtained
from the analytic solution in the small $\epsilon$ limit. In this
limit, the correlation arises from the correlation of zeroth and
first order quantities in $\epsilon$,
$\langle\bm{v}\times\bm{b}\rangle=\langle\bm{v}_o\times\bm{b}_1+\bm{v}_1\times\bm{b}_0\rangle$.
Fig~\ref{ubcorr} shows the four involved poloidal fields for the
most unstable channel mode ($\beta=100$, $\epsilon=0.3$). Direct
inspection indeed shows that the relevant fields do exhibit a
significant level of correlation.

Having gained this understanding of the transport due to the
linear modes of the instability, we now turn to the full nonlinear
problem.

\section{Numerical results}

We have simulated three different configurations, which combine a
mean vertical and/or azimuthal field with a varying field
component in the vertical or azimuthal direction, the direction of
variation being either vertical or radial. This allows us to
characterize a number of components of the turbulent resistivity
tensor.

To classify our numerical results, we use the following naming
convention: the first letter indicates the direction of the mean
field, the second letter indicates the direction of the varying
field and the third letter corresponds to the direction of the
spatial dependency. We finally add a number to identify the
simulations done in the same configuration but with different
field intensity. For instance, run XZY7 have a mean field in the
$x$ direction, a varying field in the $z$ direction with a spatial
dependency in the $y$ direction.

The various configurations and results are presented in the
corresponding subsections. They are then compared in the
concluding subsection.

\subsection{Vertical field with radial dependency $B_z(y)$}

In this case, we force the following structure for the mean magnetic field:

\begin{equation}
\langle B_z\rangle=B_z^0+\delta B^0\cos \Big(\frac{2\pi y}{L_y}\Big),
\end{equation}

The mean current is then written:

 \begin{equation}
\langle \bm{J}\rangle=\langle J_x \rangle \bm{e_x}=-\delta B^0 \frac{2\pi}{L_y} \sin\Big(\frac{2\pi y}{L_y}\Big) \bm{e_x}.
 \end{equation}

\begin{figure*}
   \centering
   \includegraphics[width=0.4\linewidth]{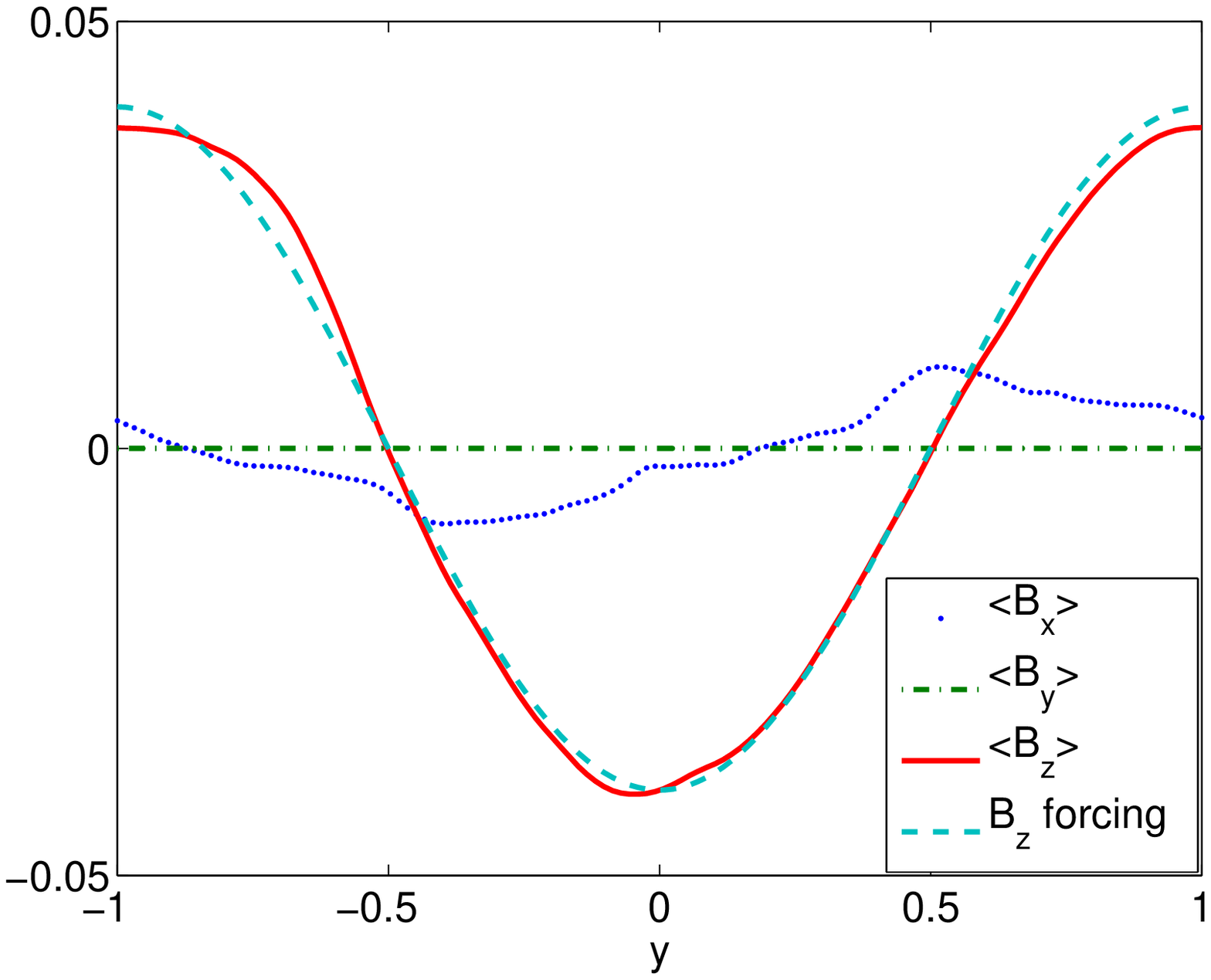}
   \includegraphics[width=0.4\linewidth]{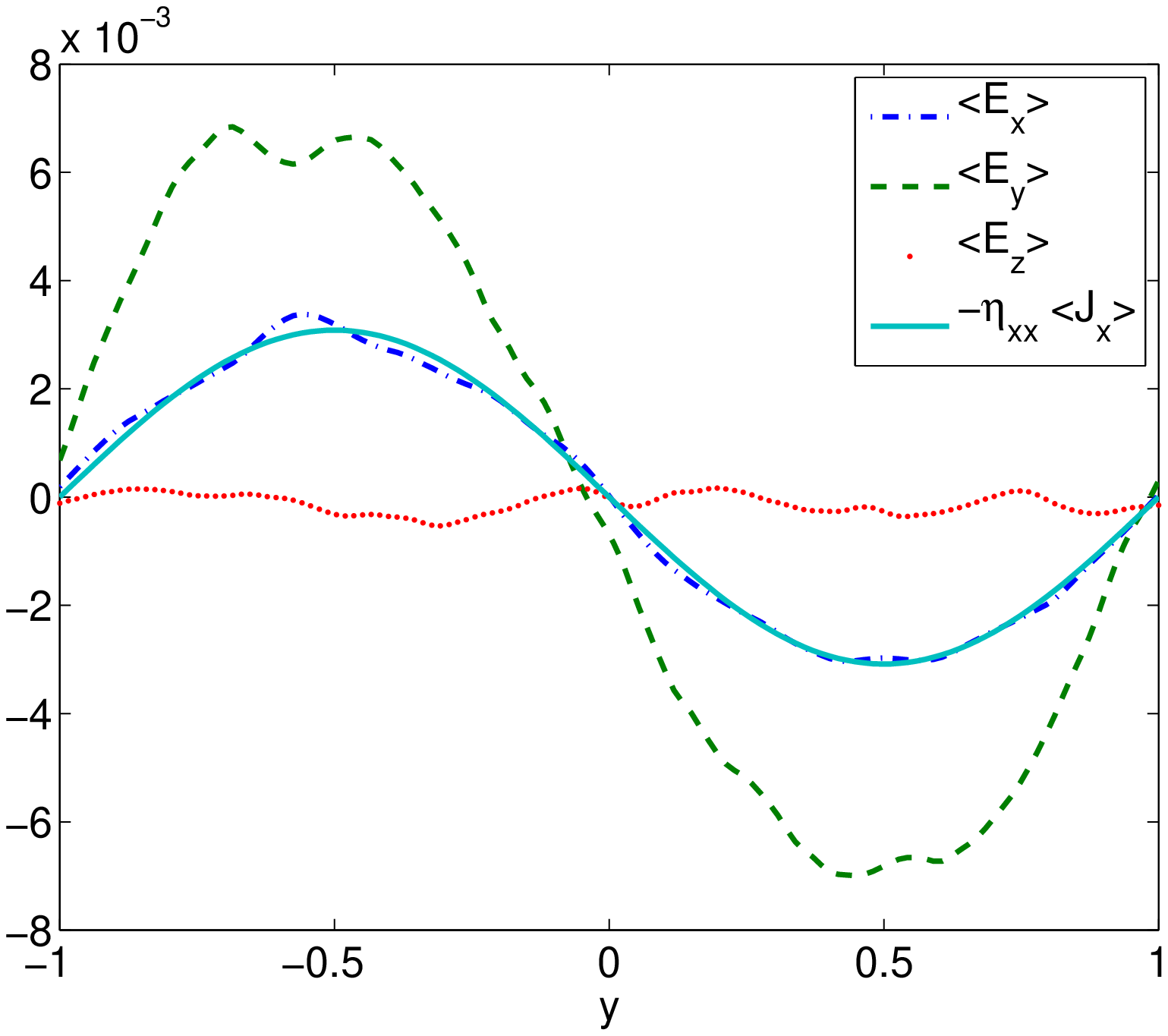}
   \caption{Mean field (left) and Emfs (right) from run ZZY4 ($B_z(y)$ case) with a mean vertical field $B_z^0=0.1$
   and $\delta B^0=0.04$. The right plot shows the best fit corresponding to the turbulent
   resistivity model. We measure in this case $\eta_{xx}=5.6\times 10^{-2}$. Note that the
   mean field $B_z^0$ has been subtracted out.}
              \label{BzyFid}%
\end{figure*}

We are therefore looking for a correlation between $J_x$ and
$\bm{\mathcal{E}}$. We show on Fig.~\ref{BzyFid} an example of a
simulation result with $B_z^0=0.1$ and $\delta B^0=0.04$
($\epsilon=0.4$). The profiles are computed from an average in
time and in the $(x,y)$ plane. From this figure, one get a
classical diagonal resitivity term of $\eta_{xx} \sim 5.6\times
10^{-2}$. We also find an non diagonal term $\eta_{yx}=8\times
10^{-2}$. We have repeated this kind of experience for various
sets of parameters as summarized in Tab.~\ref{tableBzr}.

The extraneous EMF component was already present at the linear
level, as noted in the previous section, but with the opposite
sign. We have not been able to understand the significance of this
sign difference. In our simulations, this component plays no
physical role (its rotational vanishes). This may not be a generic
feature, in particular in more complex stratified settings. This
component shows anyway that the turbulent state is anisotropic.

\begin{table}
\centering
    \begin{tabular}{| l | c | c | c | c | c |}
    \hline
    model & $B_z^0$ & $\delta B^0$ & $\eta_{xx}$ & $\eta_{yx}$ & $\alpha$ \\
    \hline
    \hline
    ZZY1       & 0.1    & 0.01   & $3.6\times 10^{-2}$    &    $2.9\times 10^{-2}$    &    $1.5\times 10^{-1}$\\
    ZZY2       & 0.1    & 0.02   & $3.7\times 10^{-2}$    &    $8.5\times 10^{-2}$    &    $1.5\times 10^{-1}$\\
    ZZY3       & 0.1    & 0.03   & $3.6\times 10^{-2}$    &    $6.8\times 10^{-2}$    &    $1.5\times 10^{-1}$\\
    ZZY4       & 0.1    & 0.04   & $2.4\times 10^{-2}$    &    $5.6\times 10^{-2}$    &    $1.3\times 10^{-1}$\\
    ZZY5       & 0.1    & 0.08   & $1.1\times 10^{-2}$    &    $1.1\times 10^{-2}$    &    $7.8\times 10^{-2}$\\
    \hline
    ZZY6       & 0.025  & 0.0025 & $4.1\times 10^{-3}$   &    $2.6\times 10^{-2}$    &    $3.7\times 10^{-2}$\\
    ZZY7       & 0.025  & 0.005   & $1.2\times 10^{-2}$   &    $-1.2\times 10^{-2}$   &    $3.3\times 10^{-2}$\\
    ZZY8       & 0.025  & 0.0075 & $7.0\times 10^{-3}$   &    $2.8\times 10^{-2}$   &     $3.7\times 10^{-2}$\\
    ZZY9       & 0.025  & 0.01     &  $5.1\times 10^{-3}$   &   $2.4\times 10^{-2}$   &      $3.5\times 10^{-2}$\\
    ZZY10     & 0.025  & 0.015   & $8.4\times 10^{-3}$   &   $1.5\times 10^{-2}$   &      $3.9\times 10^{-2}$\\
    ZZY11     & 0.025  & 0.02     & $8.5\times 10^{-3}$   &   $1.5\times 10^{-2}$   &      $3.3\times 10^{-2}$\\
    \hline

       \end{tabular}

\caption{Main results from the $B_z(y)$ (vertical field with a
radial dependency) case in the presence of a mean vertical field
$B_z^0$.}
         \label{tableBzr}
\end{table}

One may note a decrease of the turbulence efficiency in models
ZZY4-ZZY5, which may be due to the fact that increasing $\delta
B_0$ to high values leads to strong modification of the background
field. Since $B_z^0=0.1$ corresponds to the maximum of the growth
rate for the $k_z=2\pi/L_z$ mode, increasing $\delta B^0$ always
weakens the instability, which might explain these results. As
expected, this effect is not observed in $B_z^0=0.025$ models, for
which this explanation doesn't hold.

According to these results, and anticipating on the result of the
last subsection where the correlations are compared in detail, we
point out that, on average

\begin{equation}
\eta_{xx}\sim 0.235 \alpha
\end{equation}

The resulting ratio $\eta/\nu$ ($\nu=\alpha$ with our choice of
units) is substantially smaller than its linear counterpart, and
consistent with the argument exposed in the previous section.

\subsection{Toroidal field with a radial dependency $B_x(y)$}
 \begin{figure*}
   \centering
   \includegraphics[width=0.4\linewidth]{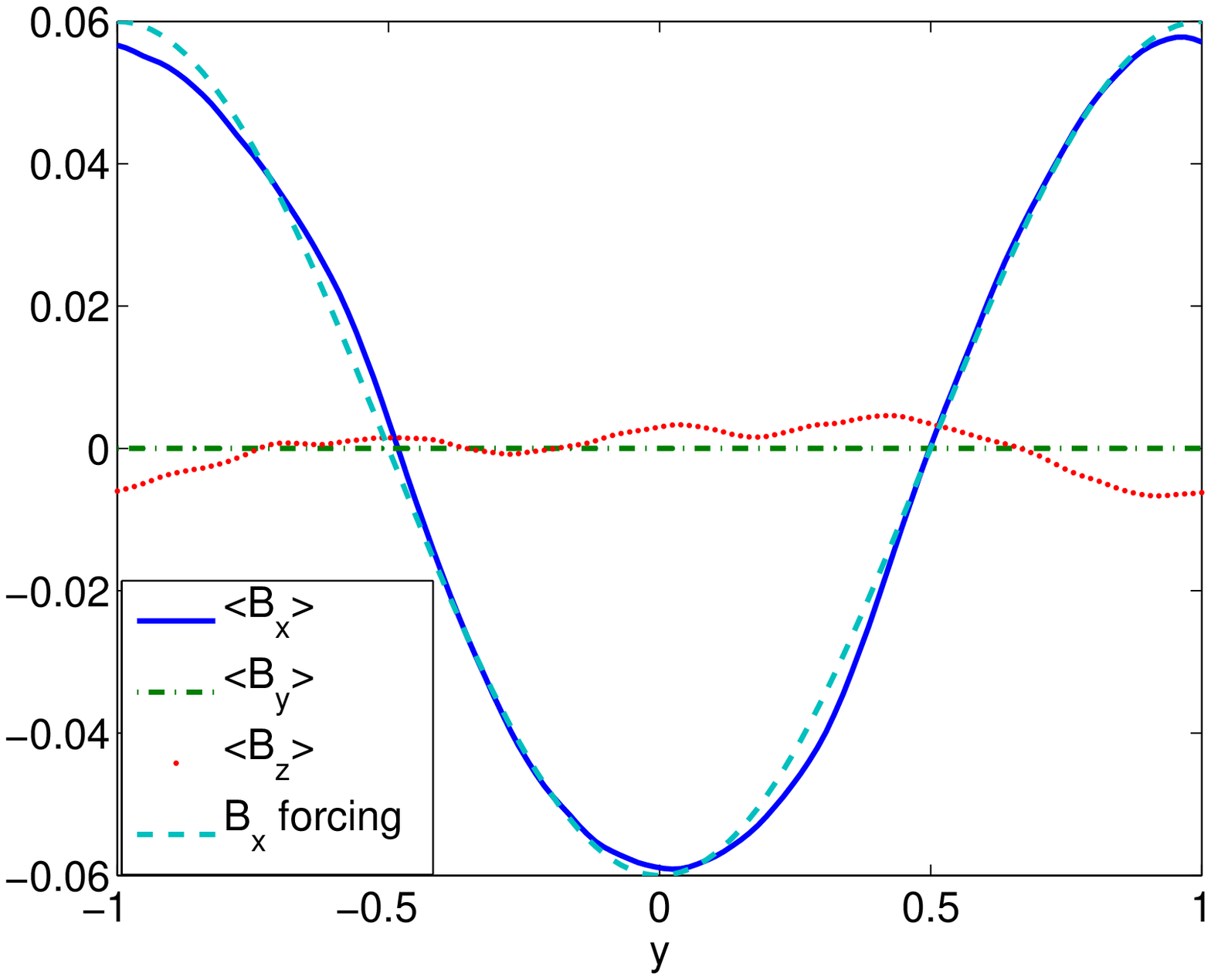}
   \includegraphics[width=0.4\linewidth]{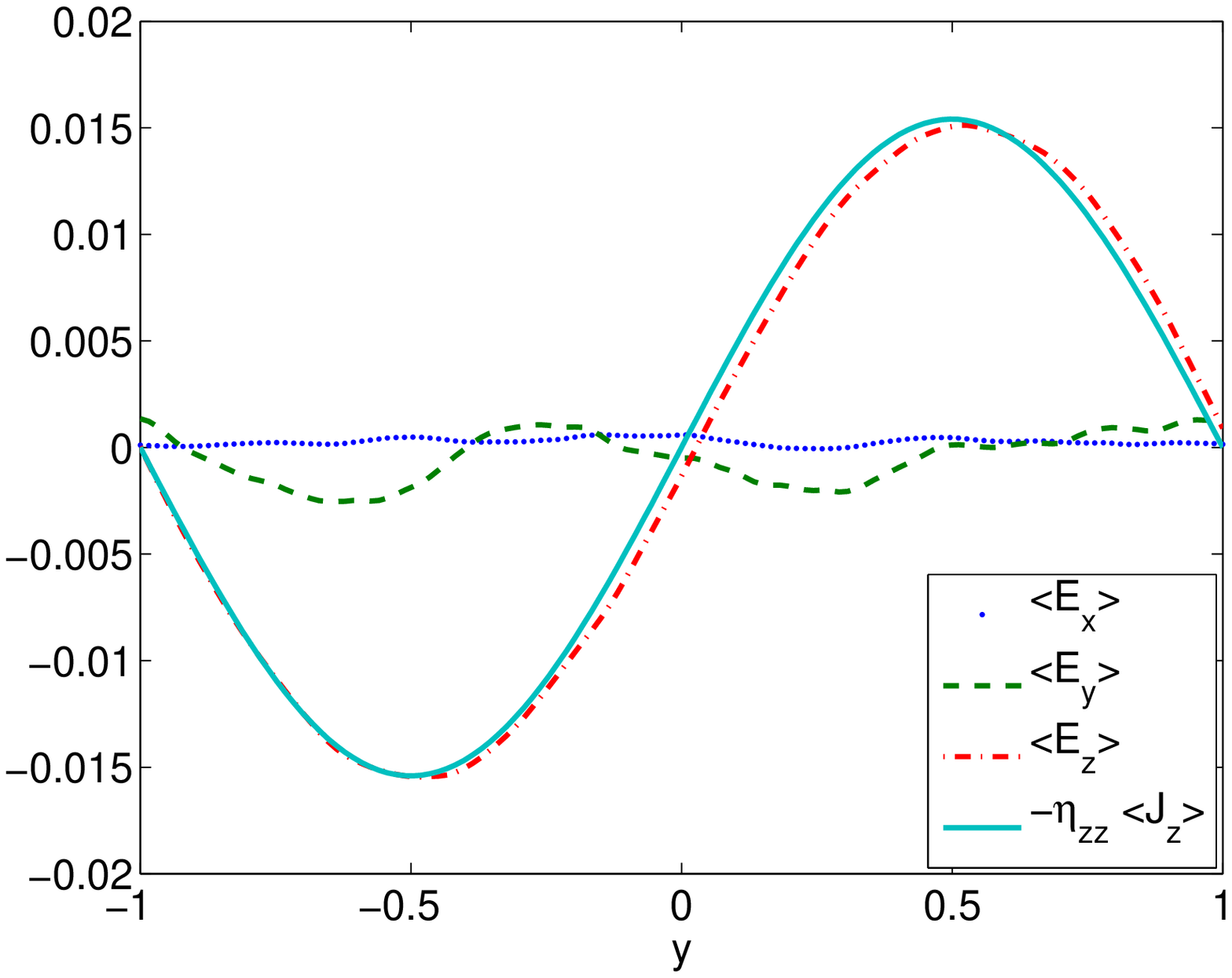}
   \caption{Mean field (left) and emfs (right) from run ZXY5 ($B_x(y)$ case) with
   a mean vertical field $B_z^0=0.1$ and $\delta B_0=0.06$. The right plot shows
   the best fit corresponding to the turbulent resistivity
   model. We find in this case $\eta_{zz}=8.2\times 10^{-2}$.}
              \label{BxyFid}%
    \end{figure*}

In this case, we impose the following field:

\begin{eqnarray}
\langle B_z\rangle&=&B_z^0,\\
\langle B_x\rangle&=&B_x^0+\delta B^0\cos \Big(\frac{2\pi y}{L_y}\Big)
\end{eqnarray}

The mean current is then defined by:

\begin{equation}
\langle \bm{J} \rangle=\langle J_z \rangle \bm{e_z}=\delta
B^0\frac{2\pi}{L_t}\sin \Big(\frac{2\pi y}{L_y}\Big)
\end{equation}

As in the previous case, we look for a correlation between $J_z$
and $\bm{\mathcal{E}}$. We show on Fig.~\ref{BxyFid} an example of
such a correlation for $B_z^0=0.1$, $B_x^0=0$ and $\delta
B^0=0.02$, from which we get $\eta_{zz}\sim 0.08$. Note that the
other components of the EMF are poorly correlated with the imposed
large scale field, indicating that either no correlation exists or
that the turbulent viscosity model does not capture its physics;
as a consequence, we have not quantified off-diagonal components
of the resistivity tensor. As previously, we reproduce this kind
of run for a range of parameters (see Tab.~\ref{tableBxr}).

\begin{table}
\centering
    \begin{tabular}{| l | c | c | c | c | c |}
    \hline
    model & $B_z^0$ & $B_x^0$ & $\delta B^0$ & $\eta_{zz}$  & $\alpha$ \\
    \hline
    \hline
    ZXY1      & 0.1   &    0.0        & 0.01   & $9.0\times 10^{-2}$    &    $1.4\times 10^{-1}$\\
    ZXY2      & 0.1   &    0.0        & 0.02   & $8.6\times 10^{-2}$    &    $1.6\times 10^{-1}$\\
    ZXY3      & 0.1   &    0.0        & 0.03   & $8.8\times 10^{-2}$    &    $1.5\times 10^{-1}$\\
    ZXY4      & 0.1   &    0.0        & 0.04   & $8.8\times 10^{-2}$    &    $1.4\times 10^{-1}$\\
    ZXY5      & 0.1   &    0.0        & 0.06   & $8.2\times 10^{-2}$    &    $1.6\times 10^{-1}$\\
    ZXY6      & 0.1   &    0.0        & 0.08   & $8.0\times 10^{-2}$    &    $1.5\times 10^{-1}$\\
    \hline
    XXY1      & 0.0    &    0.1        & 0.01  & $7.4\times 10^{-3}$    &    $7.6\times 10^{-3}$\\
    XXY2      & 0.0    &    0.1        & 0.02  & $1.6\times 10^{-2}$    &    $8.1\times 10^{-3}$\\
    XXY3      & 0.0    &    0.1        & 0.03  & $1.4\times 10^{-2}$    &    $9.1\times 10^{-3}$ \\
    XXY4      & 0.0    &    0.1        & 0.04  & $1.4\times 10^{-2}$    &    $9.2\times 10^{-3}$\\
    XXY5      & 0.0    &    0.1        & 0.06  & $1.3\times 10^{-2}$    &    $8.9\times 10^{-3}$\\
    XXY6      & 0.0    &    0.1        & 0.08  & $2.0\times 10^{-2}$    &    $1.3\times 10^{-2}$\\
    XXY7      & 0.0    &    0.1        & 0.10  & $2.1\times 10^{-2}$    &    $1.6\times 10^{-3}$\\
    \hline
       \end{tabular}

\caption{Main results from the $B_x(y)$ (toroidal field with a
radial dependency) case for several toroidal and vertical mean
fields $B_j^0$.} \label{tableBxr}
\end{table}

Comparing turbulent transport coefficient turbulent resistivity with the $B_z(y)$ case,
we note that no turbulence weakening occurs in the present case, consistently with the linear
justification (the expected linear growth rate is not modified by a varying toroidal field).
Although other experiments for larger values of
$\beta$ would allow a better determination of the resistivity, we
can still approximate

\begin{equation}
\eta_{zz}\sim 0.57 \alpha
\end{equation}

\noindent with a mean vertical field and

\begin{equation}
\eta_{zz}\sim 1.7 \alpha
\end{equation}
with a mean toroidal field.

\subsection{Toroidal field with a vertical dependency $B_x(z)$}

We define the mean field by:

\begin{eqnarray}
\langle B_z\rangle&=&B_z^0,\\
\langle B_x\rangle&=&B_x^0+\delta B^0\cos \Big(\frac{2\pi z}{L_z}\Big).
\end{eqnarray}

The mean current is then found to be:

\begin{equation}
\bm{J}=J_y\bm{e_z}=\partial_z B_x
\end{equation}
\begin{equation}
\langle \bm{J} \rangle=\langle J_y \rangle \bm{e_y}=-\delta B^0\frac{2\pi}{L_z}\sin \Big(\frac{2\pi z}{L_z}\Big)
\end{equation}

 \begin{figure*}[t!]
   \centering
   \includegraphics[width=0.4\linewidth]{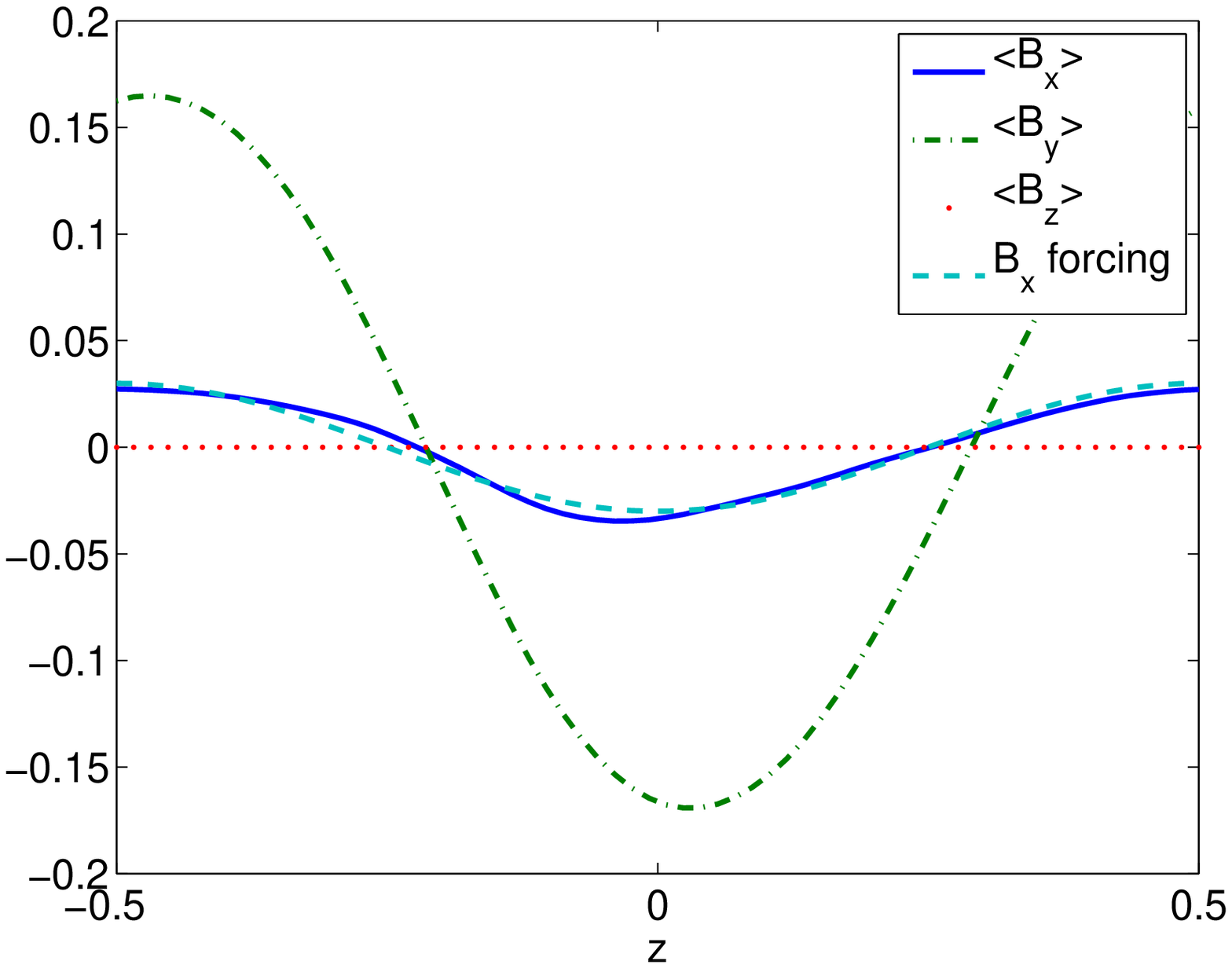}
   \includegraphics[width=0.4\linewidth]{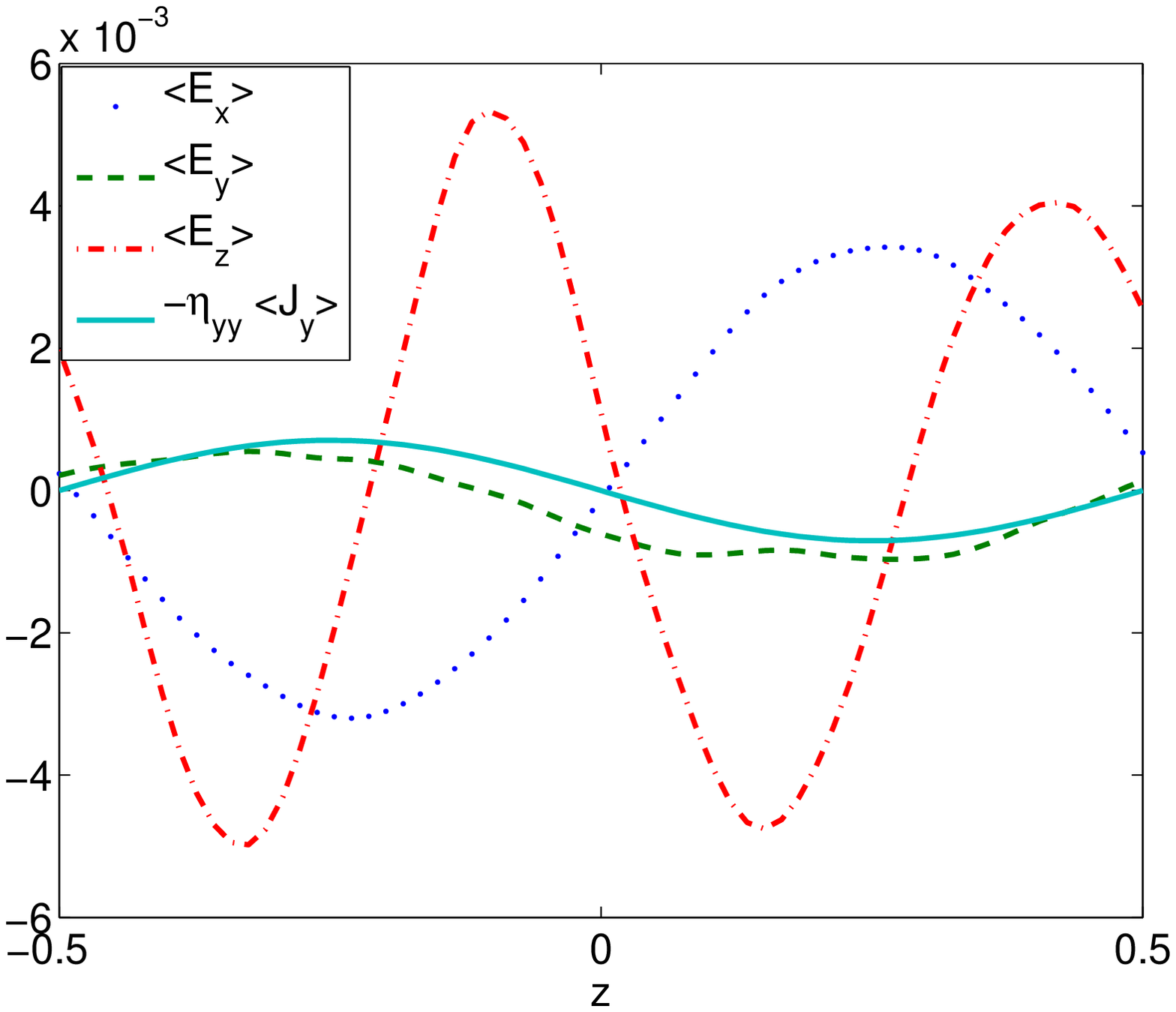}
   \caption{Mean field (left) and emfs (right) for a case with $B_z^0=0.1$ and
   $\delta B^0=0.03$ and varying $B_x(z)$.}
              \label{BxzFidPyl}%
\end{figure*}

Looking for a correlation between $J_y$ and $\bm{\mathcal{E}}$ is
more problematic in this case. Indeed, if $B_z^0\ne 0$, the
imposed $B_x(z)$ excites a channel flow solution, leading to the
production of a large scale $B_y(z)$ (see Fig.~\ref{BxzFidPyl}).
The existence of a persistent channel flow solution also leads to
a double period $\mathcal{E}_z(z)$, as observed in the simulation.




These complications induce the existence of off-diagonal
components of the resistivity tensor, which cannot be quantified
with the simple method adopted in this paper. Nevertheless, in
spite of these difficulties, the procedure used in the previous
subsections yields reasonable results for the diagonal component,
as can be seen on Fig.~\ref{BxzFidPyl}:

\begin{equation}
\eta_{yy} \sim 0.08\alpha.
\end{equation}

\noindent This value of $\eta$ is more uncertain than the other
ones derived in this work, on the order of $30\%$ to $50\%$. The
correlation with $\alpha$ is also not quite satisfied (see next
subsection), but remains acceptable at the level of precision of
determination of this quantity.

\begin{table}
\centering
    \begin{tabular}{| l| c | c | c | c | c |}
    \hline
    model & $B_z^0$ & $\delta B^0$ & $\eta_{yy}$ & $\alpha$\\
    \hline
    \hline
    ZXZ1      & 0.1    & 0.01             &  $4.8\times 10^{-3}$   &  $4.4\times 10^{-2}$\\
    ZXZ2      & 0.1    & 0.02             &  $2.5\times 10^{-3}$   &  $4.3\times 10^{-2}$\\
    ZXZ3      & 0.1    & 0.03             &  $3.7\times 10^{-3}$   &  $4.6\times 10^{-2}$\\
    ZXZ4      & 0.1    & 0.04             &  $4.8\times 10^{-3}$   &  $4.8\times 10^{-2}$\\
    ZXZ5      & 0.1    & 0.06             &  $6.8\times 10^{-3}$   &  $5.1 \times 10^{-2}$\\
    ZXZ6      & 0.1    & 0.08             &  $3.2\times 10^{-3}$    &  $7.2 \times 10^{-2}$\\
    \hline
       \end{tabular}

\caption{Main results from the $B_x(z)$ (toroidal field with a
vertical dependency) case with a mean vertical field  $B_z^0$.}
\label{tableBzxr}
\end{table}

\begin{figure*}[t!]
   \centering
   \includegraphics[width=0.4\linewidth]{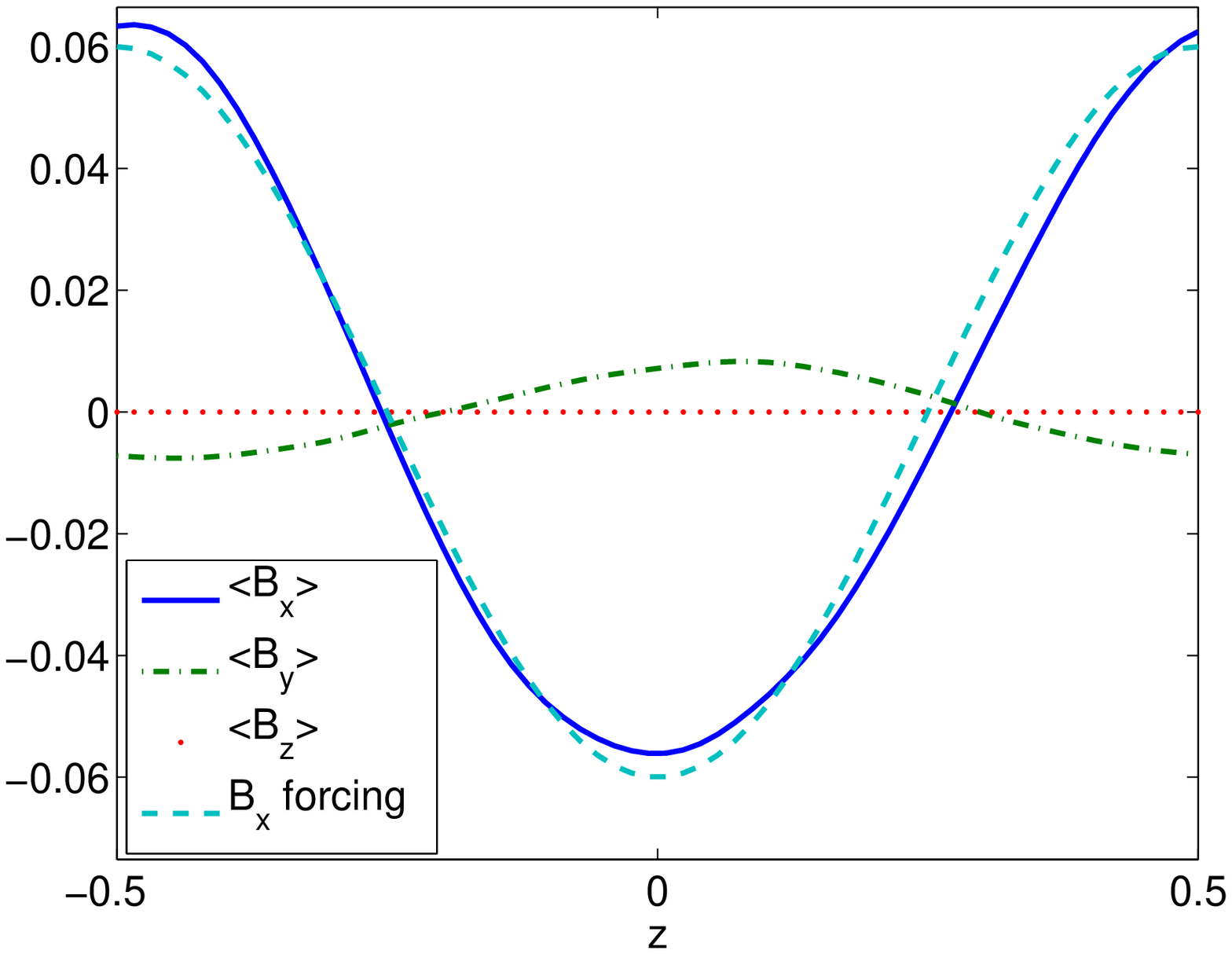}
   \includegraphics[width=0.4\linewidth]{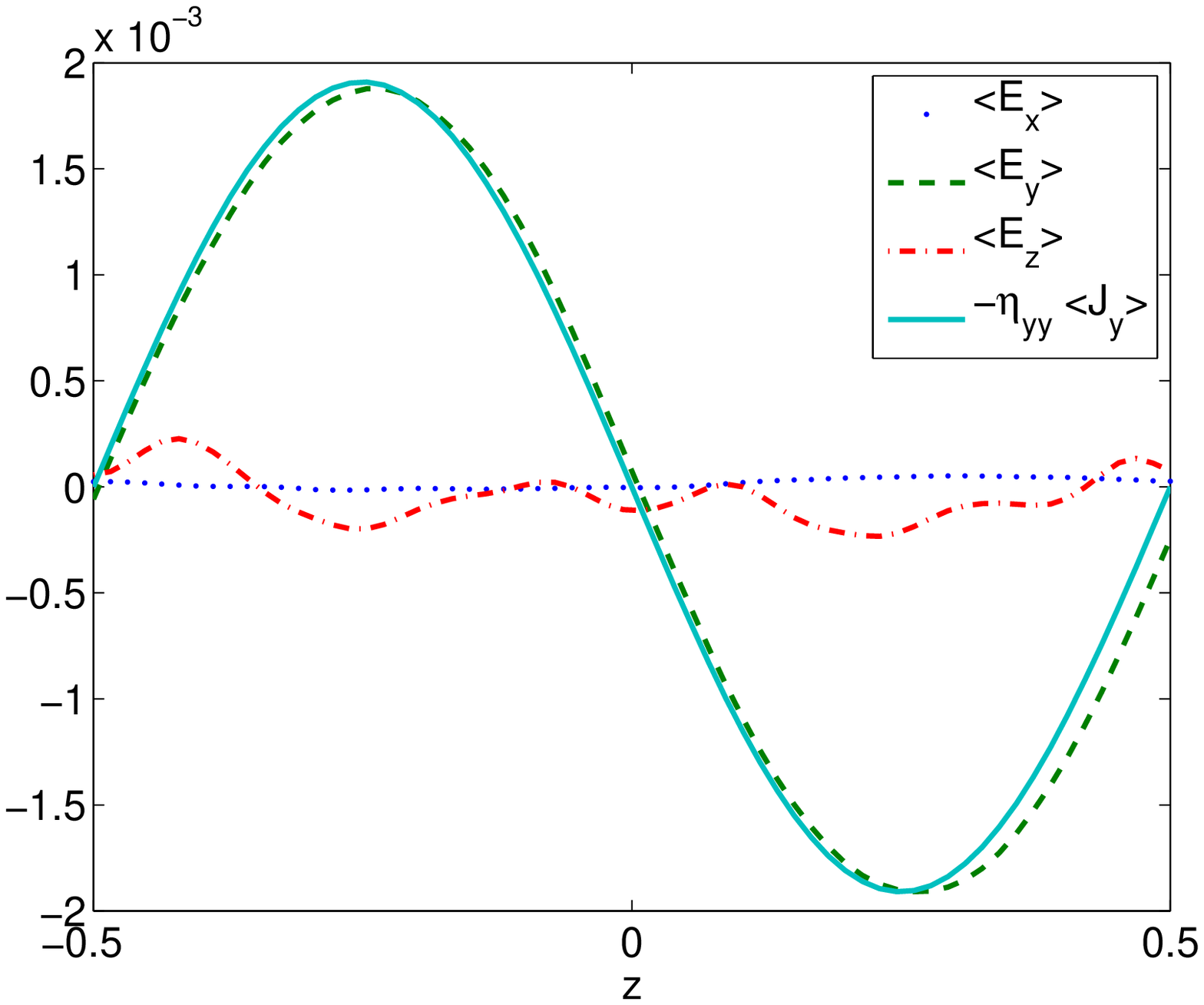}
   \caption{Mean field (left) and emfs (right) for a case with a mean toroidal flux $B_x^0=0.1$, $B_z^0=0$, $\delta B^0=0.06$ and varying $B_x(z)$. We deduce from this plot $\eta_{yy}=5.1^10^{-3}$}
              \label{BxzFid}%
\end{figure*}

The value of $\eta_{yy}$ is more precisely derived when $B_z^0=0$
which prevent channel flow formation. We present on
Fig.~\ref{BxzFid} an example of the profiles obtained from such a
simulation and the resistivity values for several runs in
Tab.~\ref{tableBxz}.

\begin{table}
\centering
    \begin{tabular}{| l | c | c | c | c | c |}
    \hline
    model & $B_x^0$ & $B_z^0$ & $\delta B^0$ & $\eta_{yy}$  & $\alpha$ \\
    \hline
    \hline
    XXZ1      & 0.1   & 0.0 & 0.01   & $5.1\times 10^{-3}$    &    $6.6\times 10^{-3}$\\
    XXZ2      & 0.1   & 0.0 & 0.02   & $5.0\times 10^{-3}$    &    $7.8\times 10^{-3}$\\
    XXZ3      & 0.1   & 0.0 & 0.03   & $5.6\times 10^{-3}$    &    $7.6\times 10^{-3}$\\
    XXZ4      & 0.1   & 0.0 & 0.04   & $5.5\times 10^{-3}$    &    $8.1\times 10^{-3}$\\
    XXZ5      & 0.1   & 0.0 & 0.06   & $5.1\times 10^{-3}$    &    $9.6\times 10^{-3}$\\
    XXZ6      & 0.1   & 0.0 & 0.08   & $6.2\times 10^{-3}$    &    $1.2\times 10^{-2}$\\
    XXZ7      & 0.1   & 0.0 & 0.1     & $6.0\times 10^{-3}$    &    $1.3\times 10^{-2}$\\
    \hline
       \end{tabular}

\caption{Main results from the $B_x(z)$ (toroidal field with a
vertical dependency) case for various toroidal $B_x^0$. }
\label{tableBxz}
\end{table}

Interestingly, $\eta_{yy}$ seems to be constant although $\alpha$
varies significantly, contrarily to previous cases. We can still
approximate:

\begin{equation}
\eta_{yy}\sim0.72 \alpha
\end{equation}

\begin{figure*}
   \centering
   \includegraphics[scale=0.7,angle=90,origin=cB]{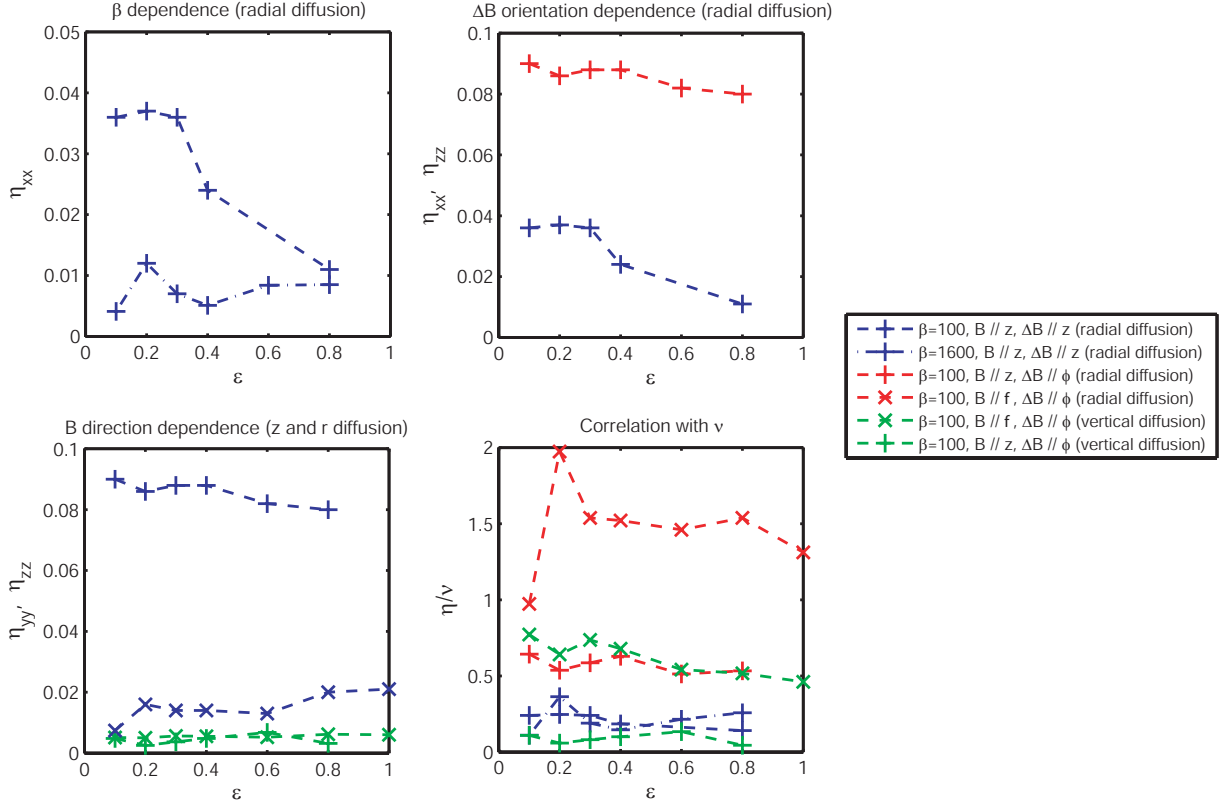}
   \caption{Dependence of the turbulent diffusivity tensor components that we were
   able to measure on $\beta$ (relative field strength) and $\epsilon$
   (relative amplitude of the varying field component), for
   various field configuration. The top left figure represents $\eta_{xx}$ for
   two different values of $\beta$. The bottom left one shows the role of the mean
   field orientation on $\eta_{zz}$. The top right compares $\eta_{xx}$ and $\eta_{zz}$.
   The bottom right figure shows the ratio of the relevant turbulent $\eta$ component over the
   turbulent viscosity. The symbols are the same as on the other subplots.}
              \label{param}%
\end{figure*}

We note however that this relation doesn't mean that the diffusive process is
very strong in the vertical direction compared to that in the horizontal direction,
as the turbulence intensity is weaker with a mean toroidal field than with a mean vertical field (compare
for instance the values of $\alpha$
in Tab.~\ref{tableBxz} with models ZXY in Tab.~\ref{tableBxr}). However, comparing only
mean toroidal field simulations (X** models), we can state that on average the
$\eta_{yy}$ coefficient is significantly smaller than
$\eta_{zz}$.

\subsection{Parameter and configuration dependence of the
turbulent resistivity tensor:}

It is of interest to compare the effect of the field configuration
and magnitude on the various turbulent resistivity tensor
components that we have characterized, most notably concerning the
radial and vertical diffusion of vertical and azimuthal field
components. To this effect, the various dependencies have been
represented on Fig.~\ref{param}.

The most significant trend is due to the direction of the mean
field. As for turbulent viscous transport (but somewhat less
dramatically), the turbulent resistive transport in the radial
direction is substantially reduced when the mean field is
azimuthal with respect to a vertical mean field, by nearly an
order of magnitude; however, and somewhat surprisingly, no such
trend is visible for the diffusion in the vertical direction
(bottom left subplot). Our simulations with a mean toroidal field
may suffer (as all others) from a lack of resolution of the
relevant small scale modes, a point that has apparently never been
adequately checked in the literature.

Another important trend, and the most critical one for the issue
of jet driving from accretion disks, concerns the relative
efficiency in the radial diffusion of vertical and azimuthal field
components (top right subplot). The diffusion of the azimuthal
field is more efficient by a factor of order 2 to 3.

There is apparently some influence of the magnitude on the mean
field on the efficiency of transport in the radial direction (top
left subplot). This trend is the same as for the turbulent viscous
transport, as shown on the bottom right subplot. The simplest
explanation for this trend is that the transport is magnetically
driven, whereas the usual adimensionalization of quantities (which
we have more or less followed in our definitions) scales transport
with the sound speed and the scale height: within factors of order
unity, $\nu = \alpha c_s H \sim \alpha \Omega H^2$ and similarly
for the various $\eta$ components. With this scaling, the various
simulations available in the literature imply that $\alpha\simeq 3
\beta^{-1/2}$. If a scaling with the Alfve\'en speed were chosen
instead, i.e., $\nu = \alpha_A v_A H$, the characteristic
dimensionless number $\alpha_A$ is independent of $\beta$ (at
least until $B_0$ is so weak that the zero mean field MRI dynamo
process takes over). It has been noted in the first subsection
that the decrease with $\epsilon$ for the largest value of $\beta$
is most likely a saturation process linked to the fact that this
choice of $\beta$ is close to the marginal stability limit of the
instability.

The last subplot (bottom right) indicates that $\beta$ (amplitude
of the mean field) and $\epsilon$ (amplitude of the field
variation) influence in much the same way the turbulent resistive
and turbulent transport. To some extent, this is also true of the
spatial orientation of the underlying quantities, and not only
their magnitudes. Overall, the turbulent resistive transport has
an efficiency which is smaller but comparable to the turbulent
viscous one, within a factor of order 2 or 3.

\section{discussion}

We have presented a systematic method to determine the turbulent
resistivity associated with MRI turbulence in accretion disks. We
have exemplified this method in the configuration radially and
vertically varying vertical magnetic fields, using nonlinear
spectral simulations of turbulence. We have also analyzed the
resistive transport due to channel modes in the linear limit, in
the presence of a radially varying vertical fields. Both the
linear calculations and the fully nonlinear simulations show that
the turbulent resistive transport is large, comparable to the
turbulent viscous transport, albeit a factor of 2 to 4 smaller.
This feature contradicts the heuristic model developed by Shu and
coworkers \citep{SGLC07} for turbulent resistive transport in YSOs
accretion disks, thereby removing an obstacle to the launching of
jets by accretion disks.

More precisely, we find that the turbulent resistivity $\eta^T$ is
largely correlated to the turbulent viscosity $\nu^T$ over a wide
range of variations of the dimensionless parameters of the problem
(but we have not explored the dependency on the dissipation
numbers, which needs an extensive study in itself). As a matter of
fact, we may define a turbulent Prandtl number
$Pm^T=\nu^T/\eta^T$, which is found to be on the order of 2 -- 5.
Second, we find that the turbulent resistivity is an anisotropic
tensor. In particular, the toroidal field ($B_x$) diffuses about 3
times more rapidly than the poloidal field ($B_z$), in the radial
as well as the vertical directions. We also find that non diagonal
terms of the turbulent resistivity tensor are non zero. As shown
by \cite{LO08a}, such terms might play an important role for disk
dynamos and large scale magnetic field generation.

It is often argued that a turbulent Prandtl number $\nu_T/\eta_T$
on the order of $R/H$ is required for turbulent disks to be able
to launch jets (see van Ballegooijen in \citealt{B89} and
\citealt{LPP94a}). In fact, because the accretion velocity in
jet-driving disks is larger than in standard accretion disks, this
requirement overestimates the necessary turbulent resistivity,
which turns out to be comparable to the turbulent viscosity in
self-consistent accretion-ejection models (see e.g.\
\citealt{CF00}; see also \citealt{RL08} and \citealt{LBR09} for a
simplified version of the same argument). This large turbulent
Prandtl number argument is also often invoked to justify that an
outer standard disk cannot transform into an inner jet-launching
one in the accretion process, but the validity of this conclusion
relies heavily on the assumed vertical structure of the models
considered (in particular the magnetic structure of the corona), a
point that has not been appropriately taken into account in the
literature up to now.

On the basis of these results, it seems quite plausible that
accretion disks have the ability to launch non stationary jets.
Although the turbulent resistivity we find is somewhat too weak to
allow for the existence of stationary accretion-ejection
structure, the anisotropy is in the right range. Nevertheless,
further work is required to get a complete characterization of the
turbulent resistivity. In particular, the correlation with
turbulent viscous transport needs to be more precisely studied, as
well as the impact of the (molecular) Prandtl number, which is
known to be strong on the momentum transport efficiency
\citep{LL07}.

While we were writing this paper, a similar study in shearing box
with the ZEUS code has appeared on the astro-ph ArXiV by Guan and
Gammie \citep{GG09}, and some discussion of the connection between
the two investigations is in order. In their paper, they impose of
mean toroidal field, wait for turbulence to reach a stationary
state. They superimpose then a sinusoidally varying field
component whose decay rate is used to quantify the turbulent
resistivity. Instead, we impose a constant sinusoidal component of
the field and study the correlation between the resulting emf and
the current in the insuing statistically stationary turbulent
regime. Guan and Gammie have also made some resolution studies,
which we have not performed, as previous experience with the
dimensionless numbers used in this work has shown us that the
dissipation scales are adequately resolved with our adopted
resolution. Finally, Guan and Gammie have looked into the effect
of the box aspect ratio, which was held fixed here. Conversely,
our parameter study is somewhat more extensive than theirs for the
type of configurations we have looked into. The majority of the
runs performed by Guan and Gammie have been made for vertical
sinusoidal component superimposed on the mean toroidal field, a
configuration we have not investigated, so that these runs are not
directly comparable to ours. Note however that using a mean
vertical field instead of a mean toroidal one is more relevant to
the question of flux diffusion for accretion-ejection models.
Nevertheless, some of the runs of \cite{GG09} have been performed
with both the mean and sinusoidal component in the azimuthal
direction, which are comparable to the ones presented here in
section 4.3. Their turbulent resistivity and viscosity values are
systematically larger than ours, but note that their $\epsilon$
parameter in these runs is four times larger than the larger one
we have used (and believe to be more relevant to disk physics). As
there is a clear trend towards a significant increase of the
turbulent resistivity and viscosity with increasing $\epsilon$ at
the larger values of $\epsilon$ shown in table 4, we conclude that
the two results are reasonably compatible. More generally
speaking, in order of magnitude, both studies are rather
complementary and agree on the fact that the turbulent resistivity
and viscosity are comparable, but runs in similar settings would
be required to make a full comparison between the two methods.

\begin{acknowledgements}

The majority of the simulations presented in this paper were
performed using the Darwin Supercomputer of the University of
Cambridge High Performance Computing Service
(http://www.hpc.cam.ac.uk/), provided by Dell Inc. using Strategic
Research Infrastructure Funding from the Higher Education Funding
Council for England. GL acknowledges support by STFC. Other
simulations have been performed on the SCCI cluster of the
\textit{Observatoire de Grenoble}.

We thank our referee, Jim Stone, for his detailed comments on this work.

\end{acknowledgements}

\bibliographystyle{aa}
\bibliography{glesur}

\end{document}